\documentclass{ifacconf}

\usepackage{graphicx,subcaption}      
\usepackage{natbib}        
\usepackage{amsmath,amsfonts,amssymb,mathrsfs,bbm} 
\allowdisplaybreaks
\usepackage{comment,soul}  
\usepackage[dvipsnames]{xcolor} 
\usepackage{algorithm,algorithmicx,algpseudocode} 
\usepackage[short]{optidef} 

\usepackage[innerleftmargin=0.2em, innerrightmargin=0.2em,skipabove=0.2em, innertopmargin=0.2em, innerbottommargin=0.2em,skipbelow=-0.05em]{mdframed}
\newmdtheoremenv[%
ntheorem = true,
]{theorem}{Theorem}
\newmdtheoremenv[%
ntheorem = true,%
]{proposition}[theorem]{Proposition}
\newmdtheoremenv[%
ntheorem = true,%
]{lemma}[theorem]{Lemma}
\newmdtheoremenv[%
ntheorem = true,%
]{corollary}[theorem]{Corollary}
\newmdtheoremenv[%
ntheorem = true, backgroundcolor=white%
]{ancillarytheorem}[theorem]{Theorem}
\newtheorem{remark}[theorem]{Remark}
\newtheorem{problem}{Problem}

\newtheorem{assumption}{Assumption}

\usepackage{tikz}
\usetikzlibrary{arrows.meta,automata}
\usetikzlibrary{shapes.geometric,backgrounds,mindmap} 
\usetikzlibrary{positioning,decorations.pathreplacing,fit,calc}
\tikzset{frame/.style={line width=2pt, inner sep=0.7em,rounded corners, draw=#1}}

\newcommand{\Depth}{1}
\newcommand{\Height}{1}
\newcommand{\Width}{1}

\newcommand{\timeset}{\mathcal{D}}
\newcommand{\Ntraj}{N_{\text{train}}}
\newcommand{\R}{\mathbb{R}}
\newcommand{\statsp}{\R^n}
\newcommand{\Sig}[3]{\mathcal{S}_{#1}^{#2}(#3)}
\newcommand{\De}[2]{\Delta_{#1}^{#2}}
\newcommand{\vectorization}[1]{\text{vec}(#1)}
\newcommand{\kron}{\otimes_K}
\newcommand{\Fit}{\mathcal{F}}
\newcommand{\TenSer}{\mathcal{T}(({\scriptstyle \R^{d+1}}))}
\newcommand{\ProdTenSer}{\circledast}
\newcommand{\shuffle}{\Cup}
\newcommand{\shuffleSet}{\mathfrak{S}(n,m)}
\DeclareMathOperator*{\argmin}{argmin}

\begin{document}
\begin{frontmatter}

\title{On the role of the signature transform in nonlinear systems and data-driven control }


\author[ETH]{Anna Scampicchio} 
\author[ETH]{Melanie N.~Zeilinger}

\address[ETH]{Institute for Dynamic Systems and Control, ETH Z\"urich, Switzerland (e-mail: \texttt{\string{ascampicc,mzeilinger\string}@ethz.ch})}

\begin{abstract}   
Classic control techniques typically rely on a model of the system's response to external inputs, which is  
difficult to obtain from first principles especially if the unknown dynamics are nonlinear. In this paper, we address this issue by presenting an approach based on the so-called signature transform, a tool that is still largely unexplored in data-driven control. We first show that the signature provides rigorous and practically effective features to represent and predict system trajectories.
Furthermore, we propose a novel use of this tool on an output-matching problem, paving the way for signature-based, data-driven predictive control.


\end{abstract}


\end{frontmatter}

\section{Introduction}
Control theory and practice are increasingly challenged to perform safe regulation of dynamical systems that are complex, nonlinear -- and unknown. Typically, control is designed by leveraging a model of the system's behavior in response to exogenous inputs. Such a model can take different forms, according to the ``palette" reviewed in~\cite{schoukens_nonlinear_2019}: moving away from the (typically unavailable) first-principles, \textit{white}-box representations, there is a wide spectrum of \textit{data-driven} methods that can be employed, depending on the amount of physical information available. When there is none, data are used to train \textit{black}-box models, which return a typically non-interpretable mechanism to predict system trajectories in presence of new inputs -- a paradigmatic example being methods based on deep learning~\citep{pillonetto_deep_2025,wen_transformers_2023}.
Due to their data-hungriness and the lack of robustness certificates, data-driven control methods have instead been privileging more interpretable models, which typically try to represent the system's behavior by means of a linear combination of suitable features. Popular and theoretically sound approaches falling into this category leverage, e.g., Koopman operator theory~\citep{mauroy_koopman_2020, bevanda_koopman_2021} and Volterra series~\citep{cheng_volterra-series-based_2017} -- however, practical rules to choose the finite-dimensional subspace of interest are still mostly heuristic and not mature yet.
\begin{mdframed}
Overall, in data-driven control there is a quest for tools to represent unknown nonlinear dynamics that are accurate without being excessively data-hungry and that can be complemented by a rigorous analysis of their performance.\end{mdframed} In this paper, we highlight the great potential of the so-called \textit{signature transform} (or, simply, \textit{signature}), a tool mostly studied in time-series analysis but largely unexplored in data-driven control, in addressing these issues. Specifically, we focus on continuous-time, input-affine dynamical systems (which capture most of the systems of interest in control -- see~\cite{isidori_nonlinear_1995}) and adapt to them the results on the signature transform, in particular stressing its \textit{universal approximation property}. Next, after reviewing the use of the signature for predicting the system trajectories corresponding to a new input, we propose a novel strategy based on the signature to tackle an output-matching, open-loop control problem.

\subsection*{Related work} \subparagraph*{The signature transform.} This mathematical object has been studied starting from the seminal work~\cite{chen_integration_1958}, and flourished in the branch of stochastic analysis known as \textit{rough path theory}~\citep{lyons_differential_2007}. Signatures found applications mostly in time-series analysis (see~\cite{fermanian_embedding_2021} and references therein), and have also been deployed in~\cite{muca_cirone_theoretical_2024} to provide a theoretical analysis of deep selective state-space models. 
\subparagraph*{Signatures in control theory.} The work in~\cite{chen_integration_1958} has been partially extended to handle input-affine dynamical systems in~\cite{fliess_fonctionnelles_1981}, but in general there has been no use of the signature in the context of data-driven control so far. The only exception is~\cite{ohnishi_signatures_2024}, a work developed independently from ours, which takes a dynamic programming viewpoint and uses the signature to parametrize the value function, not to explicitly represent system dynamics. 
\subparagraph*{Signature inversion.} When applying signatures to control (for instance, in the output-matching problem tackled in this paper), retrieving the generating signal given its signature is one of the biggest challenges. Such a problem has been theoretically investigated, e.g., in~\cite{rauscher_shortest-path_2024} and the references therein. The only work that presents a practical algorithm to tackle it is~\cite{fermanian_insertion_2023}, but the approximation level thereby obtained is typically too crude. The output-matching control approach proposed in this paper contributes to this line of research: it leverages an insight on the universal approximation property that has never been exploited so far, thus giving a new perspective on signature inversion.

\subparagraph*{(Continuous-time) system identification.} We can look at the signature-based model studied in this paper under the lens of \textit{nonlinear system identification}. In that field, most of the models for dynamical systems are expressed in discrete-time, motivated by the fact that controllers are mostly digital~\citep{schoukens_nonlinear_2019}; therefore, the literature on continuous-time identification is not as broad -- see, e.g.,~\cite[Chapter 10]{billings_nonlinear_2013} for a survey of available techniques. Yet, the systems to be modeled mostly work in continuous time: sampled-time models are typically just approximations, and they usually fail in situations in which data are irregularly sampled -- which could occur, e.g., in financial time-series or in biomedical applications. The signature-based model addresses all of these issues, considering a continuous-time representation of the dynamics and working with time-points that can be unevenly spaced.

\subsection*{Outline} The paper unfolds as follows. 
In Section~\ref{sec:problem_statement} we define the input-affine, continuous-time dynamical systems of interest, together with the adopted data-collection scheme, and we state the two main problems addressed in the paper, namely prediction and output-matching control. The signature is defined in Section~\ref{sec:setup_and_theory}, where we also present the strategy that allows us to bridge the control set-up of Section~\ref{sec:problem_statement} with the one used for the signature. 
Additionally, we review the key theoretical properties that are then deployed in Section~\ref{sec:methodology}, where the proposed approaches to tackle prediction and output-matching are presented. The numerical tests relative to these two problems are reported in Section~\ref{sec:experiments}, and Section~\ref{sec:conclusions} collects final remarks. Finally, we present in the Appendix a more detailed review of the main signature properties, culminating in the proof of the universal approximation property.

\section{Problem statement}\label{sec:problem_statement}
\paragraph*{Dynamical systems of interest.} Given a time interval $[0,T]$, consider a state vector $Z_t \colon [0,T] \rightarrow \statsp$ that evolves according to the controlled nonlinear differential equation
\begin{equation}\label{eq:controlFriendlyCODE}
    \dot{Z}_t = f_0(Z_t) + \sum_{i=1}^d f_i(Z_t)U_t^i
\end{equation}
where, for each component $i=1,\dots,d$, the signal $U_t^i \colon [0,T] \rightarrow \R$ is an exogenous input,
and the vector fields $f_0(\cdot),\dots,f_d(\cdot)$ are
$\statsp\rightarrow \statsp$ maps. They satisfy 
\begin{assumption}\label{assu:inputs_and_vectorfields}
    The exogenous inputs $U_t^i$ are integrable on $[0,T]$ for each component $i=1,\dots,d$, and the vector fields $f_0(\cdot), \dots, f_d(\cdot)$ are globally Lipschitz-continuous but \textit{unknown}.
\end{assumption}
Furthermore, for ease of exposition and to align with customary choices in control design, we make
\begin{assumption}\label{assu:piecewise_constant}
    For every $i=1,\dots, d$, inputs $U_t^i$ are piecewise-constant on every interval $[t_j, t_{j+1})$, with $\{t_j\}_{j=0}^{N}$ in the (possibly irregular) time-grid $\timeset \doteq \{ 0 = t_0 < t_1 < \dots < t_N =~ T\}$. 
\end{assumption}
\paragraph*{Data collection and problems of interest.} Given the time grid $\timeset$, we assume to collect noiseless input-state samples on $\timeset$ from $\Ntraj$ trajectories with common initial condition $Z_0 = z_0$. Note that, in this set-up, ``state" and ``output" are synonyms. 
We aim at addressing: 
\begin{mdframed}
\begin{problem}[Prediction]\label{problem:prediction}
    Given new inputs $\{\bar{U}_t^i\}_{t\in \timeset}$ with $i=1,\dots,d$, retrieve the state trajectory returned by the system~\eqref{eq:controlFriendlyCODE};
\end{problem}
\begin{problem}[Output-matching control]\label{problem:output_matching_control}
Given a desired sampled state-trajectory $\{\bar{Z}_t\}_{t\in\timeset}$, find an input which, fed into~\eqref{eq:controlFriendlyCODE}, tracks $\{\bar{Z}_t\}_{t\in\timeset}$. 
\end{problem}
\end{mdframed}

\section{Set-up and theoretical results}\label{sec:setup_and_theory}
\paragraph*{Reformulation of dynamics.} As a preliminary step in view of defining the signature transform, we rewrite~\eqref{eq:controlFriendlyCODE} in terms of the involved differentials.
For every component $i=1\dots,d$, define the signal $V_t^i = \int_0^t U_{\tau}^i d\tau$ -- which, by Assumption~\ref{assu:piecewise_constant}, will be a continuous and piecewise-linear function $t \in [0,T] \rightarrow \R$. The dynamical system~\eqref{eq:controlFriendlyCODE} can be then  re-written as
\begin{equation}\label{eq:CODE}
    dZ_t = f_0(Z_t)dt + \sum_{i=1}^d f_i(Z_t)dV_t^i \doteq f(Z_t)dX_t,
\end{equation}
where we introduced the vector representations $f(z) \doteq [f_0(z) \dots f_d(z)] \in \R^{1\times (d+1)}$ for every fixed $z \in \statsp$, and $X_t = [t \: V_t^1 \dots V_t^d]^{\top} \in \R^{(d+1) \times 1}$ for $t \in [0,T]$, which will be also named \textit{input path}. Before defining and deriving the main properties of interest of the signature of $X_t$, we first note that the solution of~\eqref{eq:CODE} (and, thus, of~\eqref{eq:controlFriendlyCODE}) is well defined:
\begin{mdframed}
\begin{prop}\label{prop:solutionCODE}
    If Assumption~\ref{assu:inputs_and_vectorfields} holds, the solution $Z_t$ of~\eqref{eq:CODE} on $[0,T]$ exists and is unique for any arbitrary, fixed initial condition $z_0 \in \statsp$. Furthermore, it is also continuous with respect to the input path $X_t$ and the initial condition $z_0$.
\end{prop}
\vspace*{-0.6em}
\begin{pf}
It is a consequence of the Picard-Lindel\"of Theorem after seeing that the input path $X_t$ is continuous and of bounded 1-variation~\cite[Definition 1.5]{lyons_differential_2007}, which follows from Assumption~\ref{assu:inputs_and_vectorfields}. See also~\cite[Theorems 3.4, 3.8, 3.18]{friz_multidimensional_2010}.
\end{pf}
\end{mdframed}
\paragraph*{Definition of (truncated) signature.} For any positive integer $k$ and an arbitrary multi-index $i_1\dots i_k$ in the set $\{0, \dots, d\}^k$, define on any sub-interval $[s,t] \subset [0,T]$ the iterated integral\footnote{Such an integral can be also inductively defined as $\Sig{s,t}{i_1i_2\dots i_k}{X} = \int_{s \leq \tau_k \leq t} \Sig{s,\tau_k}{i_1i_2\dots i_{k-1}}{X}dX^{i_k}_{\tau_k}$, starting from the line integral $\Sig{s,t}{i}{X} = \int_{s \leq \tau_1 \leq t} dX_{\tau_1}^i = X_t^i - X_s^i$ computed for $i \in \{0,1,\dots,d \}$. In the simple case in which $X_t=t$, one retrieves the classic Taylor expansion~\citep[Example II.3]{monzio_compagnoni_effectiveness_2023}.}
\begin{equation}\label{eq:singleIterInt}
    \Sig{s,t}{i_1\cdots i_k}{X} \doteq \int_{s<\tau_k < t} \dots \Big(\int_{s<\tau_1 < \tau_2} dX_{\tau_1}^{i_1}\Big) \cdots dX_{\tau_k}^{i_k}.
\end{equation}
We can group all the integrals~\eqref{eq:singleIterInt} corresponding to multi-indices of length $k$ in $k$-level tensors with canonical product $\otimes$, introducing $\Sig{s,t}{(k)}{X} \doteq \int_{s \leq \tau_1 \leq \dots \leq \tau_k \leq t} dX_{\tau_1} \otimes \dots \otimes dX_{\tau_k} \in (\R^{d+1})^{\otimes k}$. Thanks to these quantities, the \textit{signature} of the path $X$ is the infinite collection of the tensors $\Sig{s,t}{(k)}{X}$ whose elements are given by~\eqref{eq:singleIterInt}, i.e.~(see also Figure~\ref{fig:signature_visualization} for a visualization):
\begin{equation}\label{eq:signature}
    \Sig{s,t}{}{X} \doteq (1, \Sig{s,t}{(1)}{X},\Sig{s,t}{(2)}{X},\dots).
\end{equation}

\begin{figure}[h!]
\centering
\begin{tikzpicture}[framed]
    \node[scale=1] (sig) at (-0.3,0) {$\Sig{s,t}{}{X} = $};
    \begin{scope}[xshift=-1em]
    \node[scale=1] (lbrace) at (1,0) {$\Bigg($};
    \node[circle, inner sep=0.1em, fill=RedOrange,draw=yellow] (0dim) at (1.3,0) {};
    \node[] (comma1) at (1.6,-0.1) {,};
    \draw[fill=RedOrange,draw=yellow] (2, -0.5) rectangle (2.1,0.5) {};
    \node[] (comma2) at (2.3, -0.1) {,};
    \draw[fill=RedOrange,draw=yellow] (2.7, -0.5) rectangle (3.7,0.5) {};

    \node[] (comma3) at (3.9, -0.1) {,};

    \begin{scope}[xshift=13.1em,yshift=-1em] 
    \coordinate (O) at (0,0,0);
\coordinate (A) at (0,\Width,0);
\coordinate (B) at (0,\Width,\Height);
\coordinate (C) at (0,0,\Height);
\coordinate (D) at (\Depth,0,0);
\coordinate (E) at (\Depth,\Width,0);
\coordinate (F) at (\Depth,\Width,\Height);
\coordinate (G) at (\Depth,0,\Height);

\draw[yellow,fill=RedOrange] (O) -- (C) -- (G) -- (D) -- cycle;
\draw[yellow,fill=RedOrange] (O) -- (A) -- (E) -- (D) -- cycle;
\draw[yellow,fill=RedOrange] (O) -- (A) -- (B) -- (C) -- cycle;
\draw[yellow,fill=RedOrange] (D) -- (E) -- (F) -- (G) -- cycle;
\draw[yellow,fill=RedOrange] (C) -- (B) -- (F) -- (G) -- cycle;
\draw[yellow,fill=RedOrange] (A) -- (B) -- (F) -- (E) -- cycle;
\end{scope}

\node[] (comma4) at (5.85, -0.1) {,};
\node[scale=1] (dots) at (6.3,0) {$\dots$};
\node[scale=1] (rbrace) at (7,0) {$\Bigg)$};
\end{scope}
\end{tikzpicture}
\caption{Visualization of the signature represented as a collection of tensors of increasing dimension.}
\label{fig:signature_visualization}
\end{figure}
In practice, however, we will work with the \textit{truncated signature} of order $M$: 
\begin{equation}\label{eq:truncatedSignature}
    \Sig{s,t}{}{X}^M \doteq (1, \Sig{s,t}{(1)}{X},\dots, \Sig{s,t}{(M)}{X}).
\end{equation}
We now recall the properties of the signature transform that will play a pivotal role in the proposed methodology.

\paragraph*{Useful properties.}
The first notable property of the signature is the so-called \textit{Chen's identity}~\citep{chen_integration_1958}, which allows us to compute the signature of a path, e.g., on the interval $[s,t]$ knowing the signatures for the portions defined on $[s,v]$ and $[v,t]$. 
\begin{mdframed}
\begin{thm}\label{thm:chen}
    Let $X\colon [s,v] \to \R^{d+1}$ and $Y \colon [v,t] \to \R^{d+1}$ be two continuous paths, and let $X * Y$ be their \textit{concatenation}~\cite[Definition 2.8]{lyons_differential_2007}. Then, for any multi-index $i_1\dots i_k \subset \{0,\dots,d\}^k$, $\Sig{s,t}{i_1\dots i_k}{X~*~Y} = \sum_{\ell = 0}^k \Sig{s,v}{i_1\dots i_{\ell}}{X} \Sig{v,t}{i_{\ell+1}\dots i_k}{Y}$.
\end{thm}
\vspace*{-0.6em}
\begin{pf}
    It follows from Fubini's Theorem -- see~\cite[Theorem 2.9]{lyons_differential_2007}. 
\end{pf}
\end{mdframed}
The key fact that makes the signature so powerful is its \textit{universal approximation} property: roughly speaking, it states that the signature behaves like polynomials defined on the space of paths. Formally, following~\cite[Theorem 2.3]{cuchiero_expressive_2021}: 
\begin{mdframed}
    \begin{thm}\label{thm:universalApprox}
    For any time $t$ in the interval $[0,1]$, let $Z_t$ be the solution of~\eqref{eq:CODE} with initial condition $Z_0 = z_0$. There exists a \textit{time-homogeneous, linear} operator $\mathcal{L}~\colon~ \oplus_{k=0}^{M}~(\R^{d+1})^{\otimes k} \to \R^n$, depending only on $(f, M, z_0)$, such that 
    \begin{equation}\label{eq:state_representation}
        Z_t = \mathcal{L}(\Sig{0,t}{}{X}^M) + \mathcal{O}(t^{M+1}). 
    \end{equation}
    \end{thm}
    \vspace*{-0.6em}
    \begin{pf}
        See Appendix~\ref{sub:universalApproximation}.
    \end{pf}
\end{mdframed}
Therefore, one can train the linear model associated with $\mathcal{L}$ using the data-set given by the observed input-state trajectories of the system, denoted as $\{X_t(m), Z_t(m)\}_{t\in\timeset}$ with $m=1,\dots, \Ntraj$; the features are the elements of the truncated signature~\eqref{eq:truncatedSignature}. With  $\mathcal{L}$ being time-homogeneous, the vector combining the terms of the truncated signature does not change across $t$, so it can be used for the whole trajectory. Thus, once trained, the model can be deployed, e.g., to predict the system trajectory corresponding to a new input.\\
We conclude with the specific case of the signature of piecewise-linear paths.
\begin{mdframed}
\begin{prop}\label{prop:sigPiecewiseLinear}
        Consider the continuous path $X\colon [0,T] \to \R^{d+1}$, where the exogenous part is obtained by integrating the inputs $U_t^i$ satisfying Assumption~\ref{assu:piecewise_constant} for each $i=1,\dots,d$. Define the increments $\De{j}{i} \doteq X_{t_{j+1}}^i - X_{t_{j}}^i$ for $i=0,\dots, d$ and $j=0,\dots, N-1$. Then, for any multi-index $i_1\dots i_k \subset \{0,1,\dots, d\}^k$ and 
        $\{t_j\}_{j=0}^{N-1}$ in the time-grid~$\timeset$,
\vspace*{-0.8em}
\begin{equation}\label{eq:sigLinear}
    \Sig{t_j, t_{j+1}}{i_1\dots i_k}{X} = \frac{1}{k!}\prod_{\ell=1}^k \De{j}{i_\ell}.
\end{equation} 
    \end{prop}
    \vspace*{-0.6em}
    \begin{pf}
        It follows from the definition~\eqref{eq:singleIterInt} -- see also~\cite{reizenstein_iisignature_2018}.
    \end{pf}
\end{mdframed}

\begin{remark}\label{rmk:truncation}
    Note that the factorial decay in~\eqref{eq:sigLinear} -- which holds also for general paths~\citep[Proposition 2.2]{lyons_differential_2007} -- implies that low-order signature terms impact the path representation more than high-order ones. This fact can be leveraged to choose the order $M$ of signature truncation. 
\end{remark}
\section{Proposed methodology}\label{sec:methodology}
From now on, we set the state dimension $n=1$ to simplify the notation, but the results can be easily extended to the general case~\citep{monzio_compagnoni_effectiveness_2023}. We start by deriving a signature-based model of the dynamics.
\begin{mdframed}
    \begin{prop}\label{prop:useful_model}
        Consider the sampled state trajectory $\{Z_t\}_{t\in\timeset}$ obtained by feeding~\eqref{eq:CODE} with a piecewise-linear path with nodes $\{X_t\}_{t\in \timeset}$. Then, fixing the order of signature truncation $M$ and letting $L \doteq \frac{(d+1)^{M+1}-1}{d}-1$, it holds
        \begin{align}
  \begin{bmatrix}  Z_{t_1} - z_{0} \\ \vdots \\ Z_{T} - z_{0}
  \end{bmatrix} =
\underbrace{\begin{bmatrix}
       \Sig{0,t_1}{0}{X} & \cdots & \Sig{0,t_1}{d\dots d}{X}\\
        \vdots & \vdots & \vdots \\
        \Sig{0,T}{0}{X} & \cdots & \Sig{0,T}{d\dots d}{X}
  \end{bmatrix}}_{\textstyle \doteq S(X) \in \R^{N \times L}}
  \underbrace{\begin{bmatrix}
    \beta_1 \\ \vdots \\ \beta_{L} 
  \end{bmatrix}}_{\textstyle \doteq \beta } + e,  \label{eq:effLinearModel}
\end{align}
where $e \in \mathbb{R}^{N}$ lumps the residuals which, for each $j=1,\dots,N$, scale as $\mathcal{O}\Big(\frac{t_j-t_0}{t_N - t_0}\Big)^M$.
    \end{prop}
    \vspace*{-0.6em}
    \begin{pf}
        Deploying~\eqref{eq:state_representation} in Theorem~\ref{thm:universalApprox}, the state $Z_t$ at $t \in \timeset$ can be expressed by a linear combination of terms of the truncated signature $\Sig{0,t}{}{X}^M$, plus a residual. Leveraging the fact that the operator $\mathcal{L}$ in Theorem~\ref{thm:universalApprox} is time-homogeneous, we can write the model in matrix form:\begin{align}\label{eq:firstLinearModel}
  \begin{bmatrix}
    Z_{0} \\  Z_{t_1} \\ \vdots \\ Z_{T}
  \end{bmatrix} &=
\begin{bmatrix}
    1 & 0 & \cdots & 0\\
       1 & \Sig{0,t_1}{0}{X} & \cdots & \Sig{0,t_1}{d\dots d}{X}\\
       \vdots & \vdots & \vdots & \vdots \\
       1 & \Sig{0,T}{0}{X} & \cdots & \Sig{0,T}{d\dots d}{X}
  \end{bmatrix}
\begin{array}{@{}c@{}}{
  \begin{bmatrix}
    \beta_0 \\ \vphantom{\beta_1} \vdots \\ \beta_{L}
  \end{bmatrix}}\\
  \\
  \end{array} + \begin{bmatrix}
      0 \\ e_1 \\ \vdots \\ e_N
  \end{bmatrix}, 
  \notag
\end{align}
where $L = L(d+1,M)$ is the number of terms in the truncated signature of order $M$ excluding the first (always equal to 1), and is calculated as $L \doteq \sum_{k=1}^M(d+1)^k$; and $e_j$ is the residual term presented in~\eqref{eq:state_representation} . The expression in~\eqref{eq:effLinearModel} is then obtained by substituting $\beta_0 = z_0$ derived from the first equation $Z_0 = \beta_0$.
    \end{pf}
\end{mdframed}
Such a structure will now be deployed to address Problems~\ref{problem:prediction} and~\ref{problem:output_matching_control} and perform prediction and output matching.

\subsection{Problem~\ref{problem:prediction}: Prediction}
\paragraph*{Model training.} We start by using the available data-set (sampled trajectories) $\{(X_t(m)),Z_t(m)\}_{t\in \mathcal{D}}$ with $m=1,\dots, \Ntraj$ to estimate $\beta$ in~\eqref{eq:effLinearModel}. To do so, we write the model according to  Proposition~\ref{prop:sigPiecewiseLinear} for each trajectory; then, for all  $m=1,\dots,\Ntraj$, we stack all the resulting $S(X(m))$ in a matrix $\boldsymbol{S} \in \R^{N \cdot \Ntraj \times L}$. Defining $\boldsymbol{Z} \in \R^{N \cdot \Ntraj}$ as the vector stacking all the state measurements $\{(Z_{t_1}(1) - z_0), \dots,  (Z_{T}(1) - z_0), \dots, (Z_{T}(\Ntraj) - z_0)\}$, we compute the estimate for $\beta$ via linear regression: 
\begin{equation}\label{eq:linearRegressionBeta}
   \hat{\beta} = \argmin_{\beta \in \R^{L}} \| \boldsymbol{Z} - \boldsymbol{S}\beta \|^2.
\end{equation}
\begin{remark}
We can isolate a sufficient condition ensuring uniqueness of $\hat{\beta}$:
\begin{mdframed}
\begin{prop}\label{prop:uniqueBeta}
    If the $d+1$ components of the input path $X_t$ are pairwise-distinct and $L \leq N\cdot \Ntraj$, then the solution $\hat{\beta}$ in~\eqref{eq:linearRegressionBeta} is unique. 
\end{prop}
\vspace*{-0.6em}
\begin{pf}
    Let us start with $\Ntraj=1$ and $N=L$. Thanks to the uniqueness of the signature yielded by Lemma~\ref{lemma:uniqueness} (see Appendix~\ref{sub:universalApproximation}), having pairwise-distinct input components  is a sufficient condition for the (in this case, square) matrix $\boldsymbol{S}$ to be full-column rank. Such a condition holds also for $L \leq N$ or $\Ntraj > 1$.
\end{pf}
\end{mdframed}
Note that, when $L>N\cdot\Ntraj$, regularization needs to be added to retrieve a unique solution for $\beta$.
\end{remark}

\paragraph*{Prediction.} Let $\bar{U}$ be a new input sequence satisfying Assumption~\ref{assu:piecewise_constant}. After obtaining the corresponding input path $\bar{X}$ by integrating the exogenous components, we can predict the solution of~\eqref{eq:CODE} sampled at $t\in\timeset$ through the procedure summarized in Algorithm~\ref{alg:prediction}. 

\begin{algorithm}[h]
\caption{Prediction with signature model. \textit{Input}: new input path $\bar{X}$; initial state $z_0$; estimated $\hat{\beta}$ from~\eqref{eq:linearRegressionBeta}. \textit{Output}: vector containing the estimated sampled state trajectory $[Z_0 \: Z_{t_1} \, \dots \, Z_{t_N}]^{\top}$.}
\begin{algorithmic}
\State build signature matrix $S(\bar{X})$ as in~\eqref{eq:effLinearModel};
\State evaluate $\tilde{Z} = S(\bar{X})\hat{\beta}$;
\State compute $\tilde{Z} + z_0 \mathbbm{1}_{N\times 1}$ {\footnotesize (where $\mathbbm{1}_{N\times 1}$ is an $N\times 1$-vector of ones)};
\State return $[z_0 \: \tilde{Z}^{\top}]^{\top}$.
\end{algorithmic}
\label{alg:prediction}
\end{algorithm}

\subsection{Problem~\ref{problem:output_matching_control}: Output-matching control}\label{subsec:control}
We now propose the novel signature-based, open-loop control strategy to perform output matching. To enhance the tractability of the problem, we start from a reformulation of the model~\eqref{eq:effLinearModel}:
\begin{mdframed}
\begin{prop}\label{prop:model_reformulation}
        The signature-based model~\eqref{eq:effLinearModel} can be rewritten as
        \begin{equation}
  \begin{bmatrix}  Z_{t_1} - z_{0} \\ Z_{t_2} - Z_{t_1} \\ \vdots \\ Z_{t_N} - Z_{t_{N-1}}
  \end{bmatrix} \hspace{-0.2em} \approx \hspace{-0.2em}
 \left[\begin{smallmatrix}
& \Sig{0,t_1}{(M)}{X} & \\
        & \Sig{0,t_2}{(M)}{X} - \Sig{0,t_1}{(M)}{X} & \\   
        & \vdots & \\
        & \Sig{0,t_N}{(M)}{X} - \Sig{0,t_{N-1}}{(M)}{X} & \\
        \end{smallmatrix}\right]\beta \doteq \tilde{S}(X)\beta. \notag
\end{equation}
    \end{prop}
    \vspace*{-0.6em}
    \begin{pf}
        By Theorem~\ref{thm:chen}, for an arbitrary multi-index $i_1\dots i_k \subset \{0,1\dots, d\}^d$, the signature element $\Sig{0,t_{j}}{i_1\dots i_k}{X}$ is equal to 
        $\sum_{\ell = 1}^{k-1} \Sig{0,t_{j-1}}{i_1\dots i_{\ell-1}}{X} \Sig{t_{j-1}, t_{j}}{i_{\ell}\dots i_k}{X} + \Sig{t_{j-1}, t_{j}}{i_1\dots i_k}{X} + \Sig{0,t_{j-1}}{i_1\dots i_k}{X}$.
        Therefore, if $\Sig{0,t_{j}}{i_1\dots i_k}{X}$ is the entry of $S(X)$ at $(j,\tilde{k})$, we see that its last addendum is the element at $(j-1,\tilde{k})$: this motivates us to simplify the model and subtract the $(j-1)$-th row from the $j$-th for each $j=2,\dots,N$ on both sides of~\eqref{eq:effLinearModel}.
    \end{pf}
\end{mdframed}
Proposition~\ref{prop:model_reformulation} allows us to represent the model for the sampled output trajectory in terms of the one-step increments $Z_{t_j} - Z_{t_{j-1}}$ instead of the differences $Z_{t_j} - z_0$ for each $j=1,\dots,N$, and holds for general paths. We now focus on the piecewise-linear ones conforming with Assumption~\ref{assu:piecewise_constant}. By Proposition~\ref{prop:sigPiecewiseLinear} it results that the elements in matrix $\tilde{S}(X)$ are polynomials in the input increments $\De{j}{i}$, which we collect in a vector $\Delta \in \R^{N(d+1)}$. Our goal is now to find $\Delta$ such that the resulting piecewise-linear path returns the sampled state trajectory $\{\bar{Z}_t\}_{t\in \timeset}$ starting from the same initial condition $Z_0 = z_0$ used in training. Define $\bar{Z} \doteq [\bar{Z}_{t_1} - z_0,\: \bar{Z}_{t_2} - \bar{Z}_{t_1}, \: \dots, \bar{Z}_{T} - \bar{Z}_{t_{N-1}}]^{\top} \in \R^{N\times 1}$ the vector containing the sampled target state trajectory; also, denote with $\mathbb{I}_N$ the $N\times N$ identity matrix, with $\kron$ the Kronecker product and with $\vectorization{\cdot}$ the vectorization operator. Then, the output-matching problem can be formulated as the nonlinear program 
\begin{argmini}[0]{\substack{\Delta}}{\! \! \! \! \! \quad \|\bar{Z} - (\hat{\beta}^{\top} \kron \mathbb{I}_N)\vectorization{\tilde{S}(\Delta)}\|^2}{\label{eq:deltasNonlinearProg}}{}
    \addConstraint{\! \! \! \! \De{j}{0}}{= t_{j} - t_{j-1}, \qquad}{\scriptstyle j=1,\dots,N, \; \{t_j\}_{j=0}^N \in \timeset.}
\end{argmini}
Note that the constraint in~\eqref{eq:deltasNonlinearProg} is enforcing that the components $\{\De{j}{0}\}_{j=1}^N$ should be equal to the known elements in the time-grid $\timeset$.\\
The overall pipeline to address the signature-based output-matching control problem is summarized in Algorithm~\ref{alg:control}.
\begin{algorithm}[h]
\caption{Signature-based output-matching control. \textit{Input}: target sampled state-trajectory $\{\bar{Z}_t\}_{t\in\timeset}$ contained in $\bar{Z}$, estimated $\hat{\beta}$ from~\eqref{eq:linearRegressionBeta}. \textit{Output}: Exogenous piecewise-constant inputs $U_t^i$, for each $i=1,\dots,d$.}
\begin{algorithmic}
\State Obtain $\hat{\Delta}$ by solving~\eqref{eq:deltasNonlinearProg};
\State Retrieve the piecewise-linear path $X_t$;
\State Differentiate $X_t$ to obtain $U_t$.
\end{algorithmic}
\label{alg:control}
\end{algorithm}

\section{Numerical experiments}\label{sec:experiments}
We test the effectiveness of the (truncated) signature for prediction and control on dynamical systems that have the following structure:
\begin{equation}\label{eq:langevin}
    dZ_t = \theta Z_t(\mu - Z_t^2)dt + \sigma U_t dt, \quad Z_0 = 0,
\end{equation}
where $t \in [0,T]$, $U_t \colon [0,T] \to \R$, and $(\mu, \theta, \sigma) \in \R \times \R_+ \times \R_+$. Throughout all of the experiments, differential equations are integrated using the first-order Euler scheme. Exogenous inputs follow Assumption~\ref{assu:piecewise_constant}, so we will equivalently consider their integrated (piecewise-linear and continuous) version, denoted by $V_t$, and write $\sigma U_t dt = \sigma dV_t$ in~\eqref{eq:langevin}.

\subsection{Prediction task}\label{subsec:predict_test}
\begin{figure}[h!]
    \centering
    \includegraphics[width=0.5\textwidth]{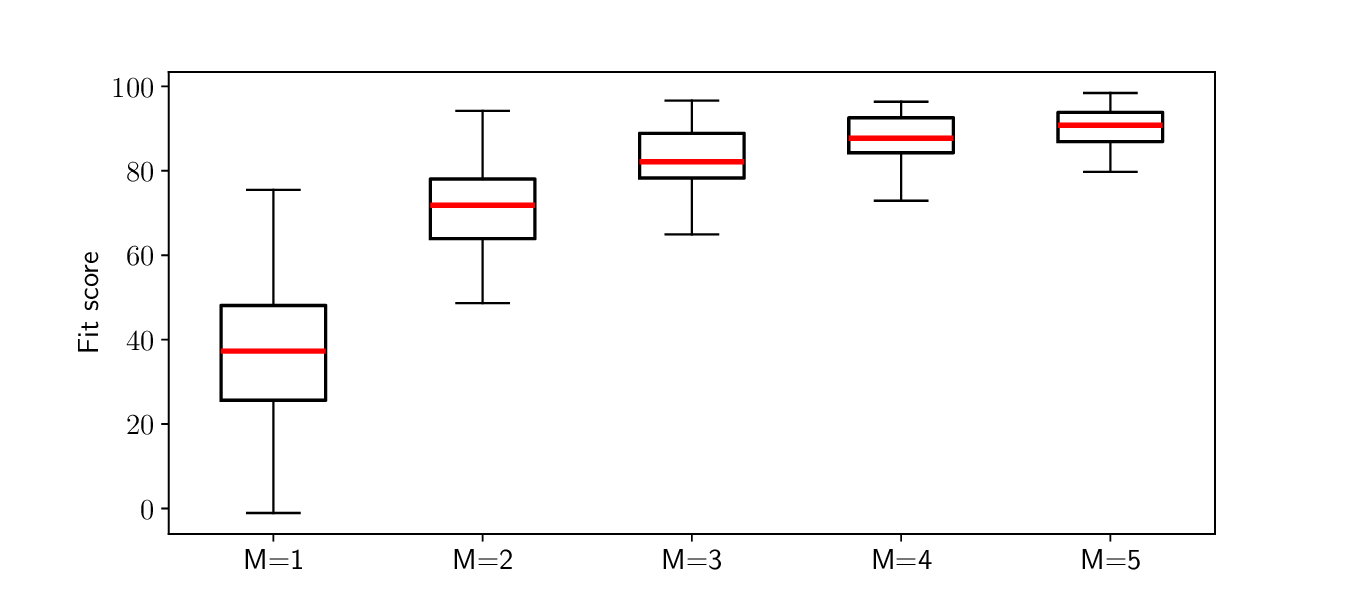}
    \caption{Values of the fit score $\Fit$ in the Monte Carlo test described in Section~\ref{subsec:predict_test} for different orders of signature truncation $M$.} 
    \label{fig:boxplot_prediction_M}
\end{figure}

We consider $N_{MC} = 50$ systems as the one presented in~\eqref{eq:langevin}, where parameters $(\mu, \theta, \sigma)$ are randomly drawn from a uniform distribution defined on $[0.5,1.5]^3$. We consider a trajectory length $T=3$ and a uniform time-grid $\mathcal{D}$ in which points are separated by an interval of length $0.01$. We train the linear regression model~\eqref{eq:linearRegressionBeta} on $\Ntraj = 40$ input-state signals, where the input $V_t$ is the piecewise-linear path whose values at $t \in \mathcal{D}$ are selected from a uniform distribution defined on $[0,5]$. The same mechanism is used to generate a new input $\bar{V}_t$ on which we test the performance in prediction. Calling ${Z}_{\text{true}}$ and $\hat{{Z}}$ the vectors in $\R^N$ containing the true state trajectory and our estimate, respectively, we quantify the performance by considering the following measure:
\begin{equation}\label{eq:fit}
    \Fit = \%100 \Bigg(1 - \frac{\| \hat{{Z}} - {Z}_{\text{true}} \|}{ \| {Z}_{\text{true}}\|}\Bigg).
\end{equation}
For each random dynamical system, we generate a new data-set and a new input trajectory, and we train the signature model using different orders of truncation $M=1, 2, \dots, 5$. The boxplot of the obtained fit scores is reported in Figure~\ref{fig:boxplot_prediction_M}, where one can notice that an almost perfect fit can be achieved already with a relatively low value of $M$. 
\vspace{-0.3em}

\subsection{Output-matching control task}\label{subsec:control_test}

\begin{figure}[h!]
    \centering
    \includegraphics[width=0.5\textwidth]{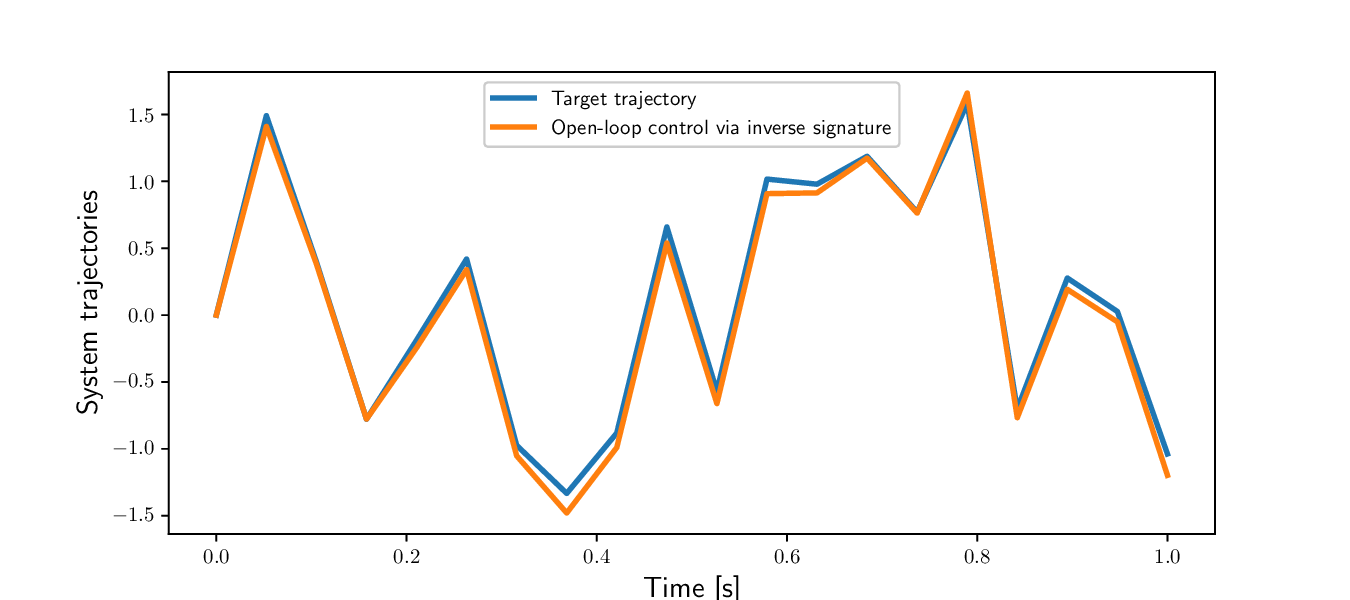}
    
    \caption{Sample performance of the output-matching strategy proposed in Section~\ref{subsec:control}. The attained fit score in this examples is $\Fit = 90.4\%$.}
    \label{fig:control_single}
\end{figure}

We now test the performance in the output-tracking problem detailed in Section~\ref{subsec:control}. We consider the system~\eqref{eq:langevin} with parameters $(\mu, \theta, \sigma) = (1,1,1)$, defined on the time interval $[0,1]$ that is regularly subdivided by a grid $\mathcal{D}$ of points equally spaced by a step of length $0.05$. The signature model with truncation order $M=4$ (leading to $L=30$) is trained with $\Ntraj = 40$ input-output trajectories, where the piecewise-linear inputs are generated in the same way as presented in Section~\ref{subsec:predict_test}, but with values in the range $[0,3]$. The same mechanism is used to generate the input which, fed into the system~\eqref{eq:langevin}, yields the desired trajectory $\bar{Z} = [0, Z_{t_1}, \dots, Z_{T}]^{\top}$ to be tracked. We carry out the procedure detailed in Section~\ref{subsec:control} and solve the nonlinear program~\eqref{eq:deltasNonlinearProg} with IPOpt implemented in CasADi~\citep{andersson_casadi_2019} and initialized on a perturbation of the true input yielding a signal-to-noise ratio of 3.5 dB. A sample outcome of the proposed procedure is presented in Figure~\ref{fig:control_single}. 

\begin{figure}[h!]
    \centering
    \includegraphics[width=0.5\textwidth]{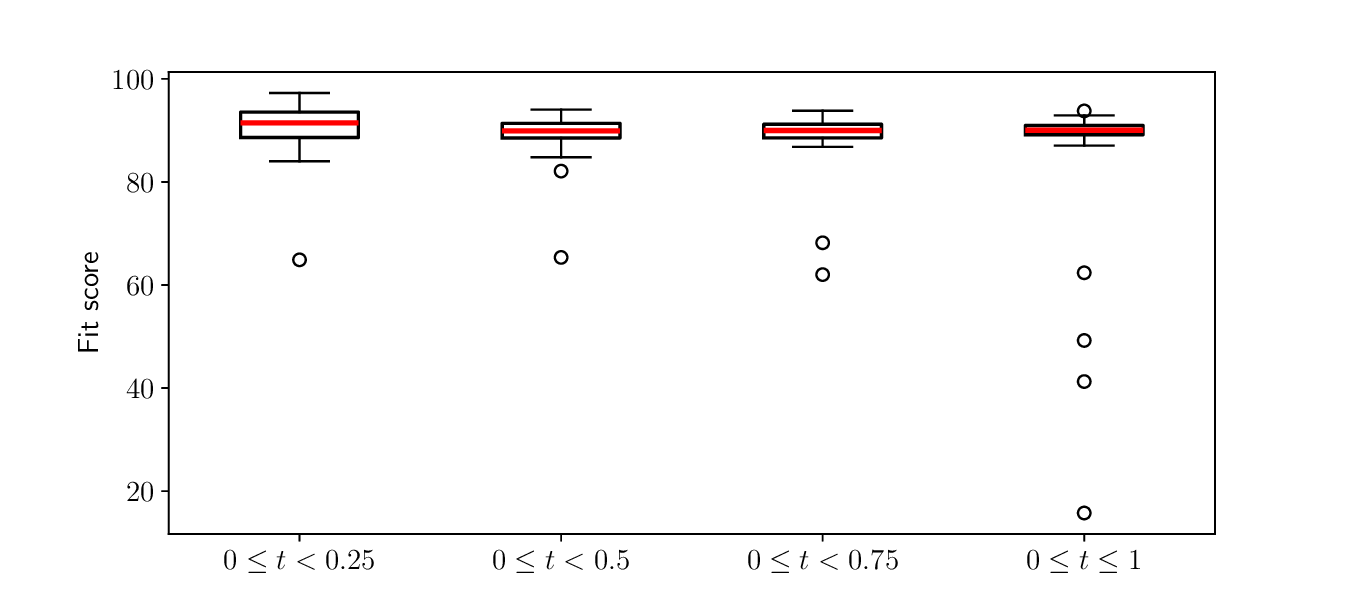}
    
    \caption{Results of the Monte Carlo test of Section~\ref{subsec:control_test}, evaluating the fit performance on trajectory portions of increasing length.}
    \label{fig:boxplot_control}
\end{figure}

We then assess the overall performance using the metric $\Fit$ defined in~\eqref{eq:fit} on a Monte Carlo study of $N_{MC}=50$ runs, randomizing over training and test data-sets, as well as on the CasADi initialization. In the boxplot of Figure~\ref{fig:boxplot_control} we display the different values of fit obtained by considering signal portions of increasing size. We can notice that the performance tends to degrade in quality as the trajectory length increases, which is an expected effect given the result of Theorem~\ref{thm:universalApprox} as well as of numerical issues in solving the nonlinear program~\eqref{eq:deltasNonlinearProg}.

\section{Conclusions and outlook}\label{sec:conclusions} 
In this paper, we showed the practical effectiveness and theoretical soundness of signature-based features in representing trajectories of unknown, input-affine (nonlinear) dynamical systems, allowing also to handle the challenging scenario in which data are irregularly sampled. Dealing with piecewise-constant exogenous inputs, we adapted the results on the signature transform to the systems of interest, and showed the performance in forecasting the system's behavior in presence of a new, unseen input. Next, we proposed a novel signature-based control for output matching, which also gives new insights in the issue of signature inversion. Investigating both facets of prediction and open-loop control opens the floor to deploy signatures in predictive control, both in the common, \textit{indirect} ``identify-then-control" scheme and in the \textit{direct} set-up reviewed in~\cite{berberich_overview_2024}. In this regard, note that the signature operates in the space of trajectories, and thus aligns with the philosophy underpinning behavioral systems theory~\citep{polderman_introduction_1998} used in direct, data-driven predictive control.\\

\subsection*{Extensions and generalization} 
\subparagraph*{Relaxing Assumptions~\ref{assu:inputs_and_vectorfields} and~\ref{assu:piecewise_constant}, and dealing with noise.} The model studied in this paper assumes global Lipschitz-continuity of the unknown vector fields: such a hypothesis can be relaxed to local Lipschitz-continuity, at the price of discussing uniqueness and continuity of~\eqref{eq:CODE} for sub-intervals of $[0,T]$. Furthermore, the choice of piecewise-constant inputs in~\eqref{eq:controlFriendlyCODE} is quite standard and also aligns with common control design practice -- however, such an assumption can be relaxed, as long as integrability still holds, and bearing in mind that one could need to numerically approximate the integral yielding the input path $X$. In general, the method presented in Section~\ref{subsec:control} can be extended to smooth input paths, possibly at the cost of increasing the number of optimization variables.
Finally, future research will deal with extending the signature-based method to the case in which measurements are noisy, deepening the preliminary results obtained in~\cite{monzio_compagnoni_effectiveness_2023}.

\subparagraph*{Implementation.} As regards the coding of the proposed control strategy, the current implementation in CasADI solves~\eqref{eq:deltasNonlinearProg} for the example presented in Section~\ref{subsec:control_test} at an average total time of 31.4 ms on a 2.6 GHz 6-Core Intel Core i7 processor. The performance can be further enhanced, both in terms of speed and solution quality, by analytically computing the gradient and the Hessian for each component of $\vectorization{\tilde{S}(\Delta)}$ with respect to $\Delta$, and including them as inputs for the optimizer. The structure of the signature matrix makes such a task feasible, and this improvement can be directly included. 
\subparagraph*{Reducing computational complexity.}
The results attained by the proposed approach are promising, thanks to the high expressiveness of the signature-based features. Yet, applying it to high-dimensional inputs could be very demanding: the truncated signature stores $\sum_{k=0}^M (d+1)^k$ terms, leading to a $\mathcal{O}((d+1)^M)$ computational complexity that could be intractable for high values of $d$ even if the truncation order is relatively small. Thus, alternative features such as the randomized signature~\citep{monzio_compagnoni_effectiveness_2023} or the log-signature~\citep{reizenstein_iisignature_2018} can be considered to address this issue and enhance the applicability of signature-based models in prediction and control.



\bibliography{references.bib}             

\appendix
\section{Review of signature properties}\label{sec:signature_properties}

In this section, we give a concise overview on the main algebraic and analytic structures of the signature, and build upon them to review the proof of its universal approximation property.

\subsection{Main algebraic structures} By its definition~\eqref{eq:signature}, the signature belongs to the space of infinite sums of $k$-level tensors denoted by $\bigoplus_{k=0}^{+\infty}(\R^{d+1})^{\otimes k}$. The latter is equivalent to the space of formal series of tensors
\begin{equation}
    \TenSer \doteq \{ \boldsymbol{a} = (a_0, a_1, \dots) | a_k \in (\R^{d+1})^{\otimes k} \: \forall k \in \mathbb{N} \}. \notag
\end{equation} 
By~\cite[Definition 2.4]{lyons_differential_2007}, $\TenSer$ is a real, non-commutative algebra. The sum of two elements $\boldsymbol{a}$ and $\boldsymbol{b}$ in $\TenSer$ is given by $(a_0+b_0, a_1+b_1, \dots)$, while their multiplication is defined as $\boldsymbol{a} \ProdTenSer \boldsymbol{b} \,  \doteq  \, (c_0, c_1, \dots)$, with $c_n$ given by the tensor products $\sum_{k=0}^n a_{k} \otimes b_{n-k}$; the unitary element for $\ProdTenSer$ is the trivial signature $\boldsymbol{1} \doteq (1, 0, \dots)$.\\ 
Focusing now on the elements~\eqref{eq:singleIterInt} of a given signature $\Sig{s,t}{}{X}$, one can prove that their product can be expressed as a linear combination of other (possibly higher-order) terms obtained through the so-called \textit{shuffle product}: 
\begin{mdframed}
\begin{prop}\label{prop:shuffle}
A permutation $\sigma$ of $\{1,\dots,n+m\}$ is called $(n,m)$-shuffle if $\sigma^{-1}(1) < \dots < \sigma^{-1}(n)$ and $\sigma^{-1}(n+1) < \dots < \sigma^{-1}(n+m)$ -- intuitively, this means that the first $n$ indices are shuffled separately from the last $m$. Define $\shuffleSet$ the collection of all of the possible $(n,m)$-shuffles of $\{1,\dots,n+m\}$, and let $I = i_1\dots i_n$ and $J = j_1\dots j_m$ be two multi-indices with $i_1,\dots,i_n, j_1,\dots,j_m$ in $\{0,\dots,d\}$. 
Considering $(r_1,\dots, r_n, r_{n+1}, \dots, r_{n+m}) \doteq (i_1,\dots, i_n, j_1, \dots, j_m)$, the \textit{shuffle product} of $I$ and $J$ is defined as $I \shuffle J \notag \doteq \{ (r_{\sigma(1)}, \dots, r_{\sigma(n+m)}) \,|\, \sigma \in \shuffleSet \}$. 
Then, it holds that, for any interval $[s,t] \subseteq [0,T]$,
\begin{equation}
    \Sig{s,t}{I}{X}\Sig{s,t}{J}{X} = \sum_{K \in I \shuffle J} \Sig{s,t}{K}{X}. \notag
\end{equation}
\end{prop}
\vspace*{-0.6em}
\begin{pf}
    We start by interpreting the integral as a linear operator and writing~\eqref{eq:singleIterInt} using the dual basis $(e_0^{\prime}, \dots, e_d^{\prime})$ of $\R^{d+1}$ as $\int_{s\leq \tau_1 \leq \dots \leq \tau_k \leq t} dX_{\tau_1}^{i_1}\dots dX_{\tau_k}^{i_k} = \langle \int_s^t dX \otimes \dots \otimes dX,\, (e_{i_1}^{\prime},\dots,e_{i_k}^{\prime}) \rangle$. The claim is then an application of Fubini's Theorem~\cite[Theorem 2.15]{lyons_differential_2007}.
\end{pf}
\end{mdframed}
From this, we have the following
\begin{mdframed}
\begin{cor}\label{cor:subalgebra}
        The terms of the signature $\Sig{s,t}{}{X}$ form a sub-algebra with respect to the shuffle product.
    \end{cor}
    \vspace{-0.6em}
    \begin{pf} Proposition~\ref{prop:shuffle} shows that the subspace of signature elements is closed with respect to multiplication.
    \end{pf}
\end{mdframed}


\subsection{Analytical properties} The starting point consists in the following result:
\begin{mdframed}
    \begin{thm}
        The signature $\Sig{0,t}{}{X} \in \TenSer$ is the solution of the differential equation 
\begin{equation}\label{eq:sigCODE}
        d\Sig{0,t}{}{X} = \Sig{0,t}{}{X} \otimes dX_t, \quad \Sig{0,0}{}{X} = \boldsymbol{1}.
    \end{equation}
    \end{thm}
    \vspace{-0.6em}
    \begin{pf}
        See~\cite[Remark 2.12]{lyons_differential_2007} and~\cite[Proposition 7.8]{friz_multidimensional_2010}
    \end{pf}
\end{mdframed}
The following key properties of the signature are its direct consequences:
\begin{mdframed}
    \begin{cor}\label{cor:propertiesSig}
    The signature (a) is invariant under time parametrization and (b) is shift-invariant. (c) Denoting with $\overleftarrow{X}_v = X_{s + t - v}$ for $v \in [s,t]$ the path $X \colon [s,t] \to \R^{d+1}$ run backwards in time, the signature satisfies $\Sig{s,t}{}{X} \ProdTenSer \Sig{s,t}{}{\overleftarrow{X}} = \boldsymbol{1}$. 
    Finally, the signature (d) exists unique, and (e) is continuous with respect to~$X$.
    \end{cor}
    \vspace{-0.6em}
    \begin{pf}
        See~\cite[Proposition 7.10]{friz_multidimensional_2010} for the proof of (a).
Claims (b)-(e) follow from the Picard-Lindel\"of Theorem stated in~\cite[Chapter 3]{friz_multidimensional_2010}. Note that (c) can also be derived from Theorem~\ref{thm:chen}.
    \end{pf}
\end{mdframed}

\subsection{Universal approximation}\label{sub:universalApproximation} 
This section presents the proof of Theorem~\ref{thm:universalApprox}, which hinges on the Stone-Weierstrass approximation Theorem:
\begin{mdframed}
    \begin{thm}\label{thm:stoneweierstrass}
        Let $\mathscr{A}$
be a compact Hausdorff space, $\mathscr{C}(\mathscr{A}, \R)$ the set of continuous real-valued functions defined on $\mathscr{A}$, and let $A$ be a sub-algebra of $\mathcal{C}(\mathscr{A},\R)$ that contains a non-zero constant function. Then $A$ is dense in $\mathcal{C}(\mathscr{A}, \R)$ if and only if it separates points.
    \end{thm}
    \vspace{-0.6em}
    \begin{pf}
        See, e.g.,~\cite[Theorem 7.32]{rudin_principles_1976}.
    \end{pf}
\end{mdframed}
We now show that the signature terms satisfy Theorem~\ref{thm:stoneweierstrass}. The space $\mathscr{A}$ is given by the compact space of continuous paths with bounded 1-variation. We showed in Corollary~\ref{cor:subalgebra} that signature elements form a sub-algebra, and the non-zero constant therein contained is the neutral element with respect to multiplication $\ProdTenSer$, which is $\mathbf{1}$. The point-separation property is ensured by the following Lemma, which is satisfied because the input path $X$ contains a time component.
\begin{mdframed}
\begin{lem}\label{lemma:uniqueness}
        If $X,\, Y\colon [s,t] \to \R^{d+1}$ have at least one strictly monotone component, then $\Sig{s,t}{}{X} \neq \Sig{s,t}{}{Y}$.
    \end{lem}
    \vspace*{-0.6em}
    \begin{pf}
        This sufficient condition can be derived from the results in~\cite{hambly_uniqueness_2010}.
    \end{pf}
\end{mdframed}
We can now apply Theorem~\ref{thm:stoneweierstrass} to the solution map of~\eqref{eq:CODE}, which by Proposition~\ref{prop:solutionCODE} is continuous in $X$, and obtain the claim of Theorem~\ref{thm:universalApprox}. 
Finally, note that the time interval $[0,1]$ in the statement is taken without loss of generality thanks to Corollary~\ref{cor:propertiesSig}(a).

\end{document}